# Low-cost and Convenient Fabrication of Polymer Micro/Nanopores with the Needle Punching Process and Their Applications in Nanofluidic Sensing

Rui Liu, [1#] Zhe Liu,[1#] Jianfeng Li,[1] and Yinghua Qiu[1,2,3]*

1. Key Laboratory of High Efficiency and Clean Mechanical Manufacture of Ministry of Education, National Demonstration Center for Experimental Mechanical Engineering Education, School of Mechanical Engineering, Shandong University, Jinan, 250061, China

2. Shenzhen Research Institute of Shandong University, Shenzhen, 518000, China

3. Suzhou Research Institute of Shandong University, Suzhou, 215123, China

# R. L. and Z. L. contributed equally.

*Corresponding author: yinghua.qiu@sdu.edu.cn

**ABSTRACT**

Solid-state micro/nanopores play an important role in the sensing field because of their high stability and controllable size. Aiming at problems of complex processes and high costs in pore manufacturing, we propose a convenient and low-cost micro/nanopore fabrication technique based on the needle punching method. The thin film is pierced by controlling the feed of a microscale tungsten needle, and the size variations of the micropore are monitored by the current feedback system. Based on the positive correlation between the micropore size and the current threshold, the size-controllable preparation of micropores is achieved. The preparation of nanopores is realized by the combination of needle punching and chemical etching. Firstly, a conical defect is prepared on the film with the tungsten needle. Then, nanopores are obtained by unilateral chemical etching of the film. Using the prepared conical micropores resistive-pulse detection of nanoparticles is performed. Significant ionic current rectification is also obtained with our conical nanopores. It is proved that the properties of micro/nanopores prepared by our method are comparable to those prepared by the track-etching method. The simple and controllable fabrication process proposed here will advance the development of low-cost micro/nanopore sensors.

**KEYWORDS:**

## I. INTRODUCTION

With the development of the micro/nano manufacturing technology,[1] various solid-state micro/nanopores have been successfully fabricated, and widely utilized in biosensing and other fields due to their controllable size and high stability.[2-7] Biosensors based on micro/nanopores can detect biomolecules or particles in situ by measuring ionic currents in solutions.[2]

The main detection principles of micro/nanopore sensors are resistive-pulse sensing (RPS),[2, 8] and steady-state current method.[5, 9] During the resistive-pulse detection, analytes are driven through solid-state pores by voltage or hydrostatic pressure, and the current signals are recorded.[10] Through analyzing the recorded current traces, the event frequency, blocking current amplitude, and duration time can be obtained, which correspond to the concentration, size, and surface charge density of analytes respectively.[11, 12] Moreover, the RPS technique can also be used to characterize other properties of analytes, such as softness[13] and deformation.[14, 15] The strong correlation between the analyte size and the micro/nanopore size ensures the high sensitivity of the resistive-pulse sensing.[8, 10] When the nanopore size is less than 2 nm, it can be used for the detection of DNA and RNA molecules.[16-19] With a pore size of ~20 nm, nanopores can be employed for the detection of protein molecules.[20, 21] Nanopores with a diameter of several hundred nanometers can be utilized for the detection of bio-nanoparticles, such as viruses and liposomes.[22, 23] With an even larger pore diameter, micropores can be used for the detection of cells and microparticles.[24-26] Since the size of analytes in resistive-pulse detection is unknown, the micro/nanopore needs to have a stable and controllable size to quantitatively obtain their physical and chemical characteristics.[10]

In the confined space of the pore, ionic transport is significantly regulated by the physical and chemical properties of pore surfaces.[27-29] The nanopore surface can be designed to interact with analytes in solution. According to the variations of ion current, the information of analytes can be revealed.[30] This is called the steady-state current method,[5, 9] which can be used for the detection and quantitative analysis of small molecules in solution, such as organic molecules and ions.[31, 32] To improve the detection accuracy, the pores are generally with sizes below 100 nm. The nanopore surface is usually modified to interact with analytes in solution in steady-state current detection.[5, 9] Through the quantitative comparison of the ion current obtained with and without nanopore modification, the presence, concentration, and type of analytes can be characterized.[5, 9] Consequently, the pore surface is preferably to have some chemical groups, such as -COOH, to facilitate the diversity of modifications.[5] In summary, high-performance micro/nanopore sensors need to meet the stability and controllability of pore sizes and surface properties.[5, 10]

The fabrication techniques of solid-state micro/nanopores include beam milling,[33] dielectric breakdown,[34] track etching,[3] and glass nanopipette pulling.[35] Li et al.[36] fabricated nanopores on $SiN_X$ films using the argon ion beam. The irradiation time is controlled by the feedback control system to realize the regulation of the nanopore size. Focused ion beam (FIB) has been widely used in the manufacture of solid-state pores.[37, 38] Because of the large size of ions, nanopores prepared by FIB are usually greater than 10 nm. To prepare nanopores with smaller sizes, high-energy electron beams emitted in TEM are used.[2, 39] Chang et al.[40] prepared sub-10 nm nanopores with TEM milling. This technology is mainly suitable for $SiN_X$, $MoS_2$, and other materials with nanometer thickness. Beam milling usually requires expensive equipment and a harsh experimental environment. However, it is widely used due to the advantages of high repeatability and high precision.[2] Dielectric breakdown, track etching, and glass nanopipette pulling technology are commonly used in a laboratory for the convenient fabrication of nanopores. Kwok et al.[41] successfully fabricated $SiN_X$ nanopores with the dielectric breakdown technique and achieved the adjustment of pore sizes by varying pH and voltages. This technique is simple to operate. However, it can only fabricate nanopores in insulating materials with a thickness of less than 30 nm, and the pore location cannot be determined. Tracked etching is mainly applicable to polymer films. Harrell et al.[42] fabricated conical nanopores using the track-etching technique and found a positive correlation between cone angles and voltages. Xiao et al.[43] prepared dumbbell-shaped nanopores by adjusting the etching temperature and the concentration of the etching solution. This technique is simple to operate, but the formation of ion tracks requires high-energy heavy-ion accelerators.[44] Glass nanopipette pulling stems from the good ductility of glass. By pulling the molten glass tube, micro/nanoscale conical pores can be created.[35] The technique is low-cost and easy to operate. While the glass nanopores have complex shapes that usually require electron microscopy to characterize.[45, 46]

To promote the practical application of micro/nanopore sensors, the fabrication process of solid-state pores is better to be convenient and at a low cost. Referring to the work by Willmott et al.[47] in which a tungsten needle was used to pierce through thermoplastic polyurethane (TPU) films to create a tunable nanopore under mechanical stretching, here we developed a simple preparation technique for polymer micro/nanopores at low-cost. With our method, micropores can be prepared with controllable sizes, which keep constant during the later application. Using the fabricated micropores, resistive-pulse detection of 400 nm particles was conducted. The blockade current amplitude shows direction-dependent which is consistent with our previous data with micropores prepared with track etching.[48] Furthermore, with the combination of needle punching and chemical etching, controllable preparation of nanopores is achieved. Our prepared nanopores have significant ionic current rectification, which indicates that the surfaces of nanopores have uniform carboxyl groups.[3, 26, 49] In this work, the performance of micro/nanopores prepared by the needle punching method is comparable to that prepared by the track-etching technique.[25, 48, 50] Taking advantage of the convenient and low-cost fabrication process, our method will promote the development and application of micro/nanopore sensors.

## II. EXPERIMENTAL METHODS AND DETAILS

Material: The polyethylene terephthalate (PET) film with a thickness of 12.5 μm used was purchased from Dongxuan Plastic Products Co., Ltd (Suzhou, China). The tungsten wire with a diameter of 0.4 mm was purchased from Hebei Zhanmo Metal Materials Company (Hebei, China). KCl or NaCl solutions were prepared with deionized water (18.2 MΩ, Direct-Q 3UV, MilliporeSigma, Burlington, MA, USA) and adjusted to pH=10. Chemicals used in the experiments were purchased from Sigma Aldrich unless otherwise stated.

Instruments: A 5V/8A switching power supply (S-40-5, Minganxin Technology, China) was used for the preparation of tungsten needles. An injection pump (LSP02-2A, Longer Pump, Baoding, China) was used to control the feed of tungsten needles. A current feedback program was written in Python to control LSP02-2A to automate the preparation of micro/nanopores. A Keithley 6487 (Keithley Instruments, Solon, Ohio, USA) was employed to detect ionic current. Scanning electron microscopy (JEOL JSM-7800F, Tokyo, Japan, and Carl Zeiss Gemini 300, Oberkochen, Germany) was used to image tungsten needles and micropores.

Particle detection and ion current acquisition: The particle suspension was prepared with 0.1 M KCl solution pH=10 containing Tween 80 with a volume fraction of 0.01%.[25, 26, 49] The carboxylated polystyrene particles (ACME Microspheres Inc, Indianapolis, USA) with a diameter of 400 nm were used. The particle concentration was ~$10^9$ /mL. Current traces were recorded with Axopatch 200B and Digidata 1550B (Molecular Devices Inc, USA). The sampling frequency was 20 kHz and the low-pass Bessel filtering frequency was 1 kHz.[25, 26, 49]

## III. RESULTS AND DISCUSSION

### A. Fabrication of Tungsten Needles

Tungsten needles with microscale tips were prepared by dynamic electrochemical etching.[51-53] The preparation process is shown in Figure S1a, with the tungsten wire as the anode immersed in NaOH etching solution, and the Cu wire as the cathode. The following electrochemical reactions occur at the cathode and anode, respectively.[51]

Cathode: $6H_2O + 6e^- \rightarrow 3H_2(g) + 6OH^-$

Anode: $W(s) + 8OH^- \rightarrow WO_4^{2-} + 4H_2O + 6e^-$

The surface tension of the liquid causes the tungsten wire to form a meniscus-like structure, resulting in an accelerated chemical etching rate of the tungsten wire near the liquid surface. With the lifting of the tungsten wire, the "necking" phenomenon appears near the surface of the etching solution. Eventually, the tungsten wire breaks because it cannot bear the weight of the tungsten wire below the liquid surface, and a microscale tip is formed at the fracture.[51]

In the preparation process of the tungsten needles, the tip cone angle and curvature radius are affected by several factors.[51] Therefore, we investigate the influences of immersion time, lifting speed, immersion depth, and NaOH concentration on the morphology of the prepared tungsten needles. The immersion time refers to the time that the tungsten needles stay in the NaOH solution before being lifted. The default immersion time, lifting speed, immersion depth, and NaOH concentration are 7 min, 0.55 μm/s, 3 mm, and 0.75 M, respectively. Figure S1b shows the tip cone angles and curvature radius of the tungsten needles under different immersion times, both of which decrease rapidly at first and then tend to be stable with the increase of the immersion time. The influence of the lifting speed on the tip cone angle and curvature radius of the tungsten needles is shown in Figure S1c. The lifting speed slower than 0.6 μm/s is suitable for the preparation of needles with a small tip cone angle and radius of curvature. From Figure S1d, immersion depths greater than 3 mm are more suitable for the preparation of small needle tips. The influence of NaOH concentration is shown in Figure S1e. At 0.5 M, no electrochemical etching occurs on the tungsten wire. In NaOH solutions higher than 0.5 M, the obtained tip cone angle stays near 15°. By adjusting the above parameters, we can realize the preparation of a tungsten needle with a desired tip cone angle, which can be applied for the fabrication of micro/nanopores based on the needle punching method.[54]

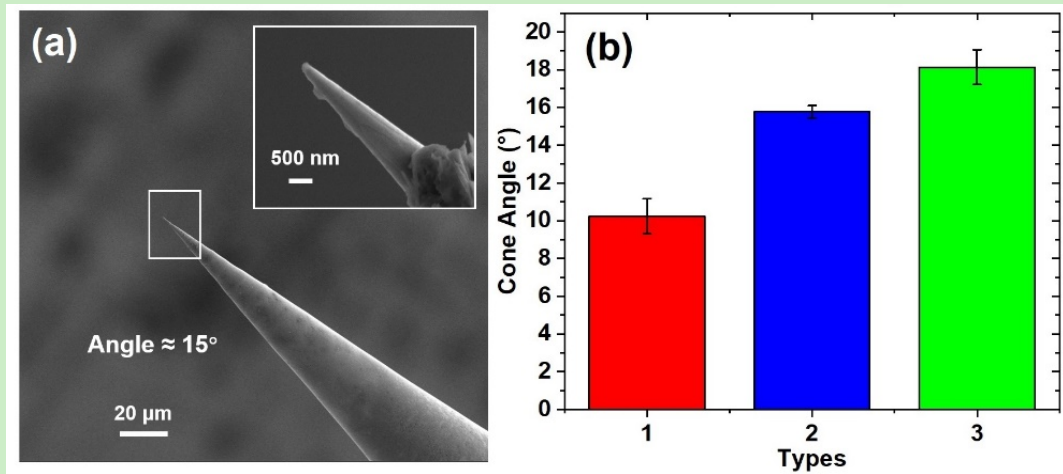

**Figure1.** Tungsten needle preparation for the micro/nanopore fabrication. (a) Scanning electron microscopy (SEM) images of tungsten needle with a cone angle of 15° and a tip diameter of ~200 nm. (b) Controllable preparation of tungsten needles with three different cone angles.

In the manufacturing process of micro/nanopores based on the needle punching method, the plastic deformation and tearing of thin films are induced by the controllable feed of the tungsten needle.[54] To ensure the strength and durability of the tungsten needle, the tip cone angle should not be too small. With a comprehensive consideration of the four factors shown in Figure S1, tungsten needles with tip cone angles of 10°, 15°, and 18° are successfully prepared (Figure 1b). Figure 1a shows a tungsten needle with a tip cone angle of 15°. The parameters used in the preparation process are: immersion time 8 mins, lifting speed 0.2 μm/s, immersion depth 2 mm, and NaOH concentration 0.75 M, respectively. From the SEM images, the prepared tungsten needles have smooth surfaces. Note that the agglomerates attached to the needle surface at the very tip may be dust or residual salt crystals left from the cleaning process. Within the first 20 μm, the linear cone boundary provides each needle with a stable cone angle. The plastic films used in this work are 12.5 μm in thickness. So we are more concerned about the angle of the first 20 μm of the tungsten needles. The tungsten needles with tip cone angles of 10° and 18° are shown in Figure S2. From the comparison among SEM images of tungsten needles with three different cone angles, the tungsten needles with a 15° and 18° tip cone angle have a better cone angle uniformity. As shown in Figure S3, the tungsten needles with different cone angles are prepared which have high controllability. The high repeatability of the tip cone angle and the surface smoothness of the tungsten needle are of great significance for the preparation of micro/nanopores with smooth inner surfaces and a desired cone. The fabricated tungsten needles are used to prepare micropores on different polymer films, such as PET, TPU, and Teflon (Figure S4). Here, we show the commonly used polymer film PET as an example.

## B. Fabrication of Micropores

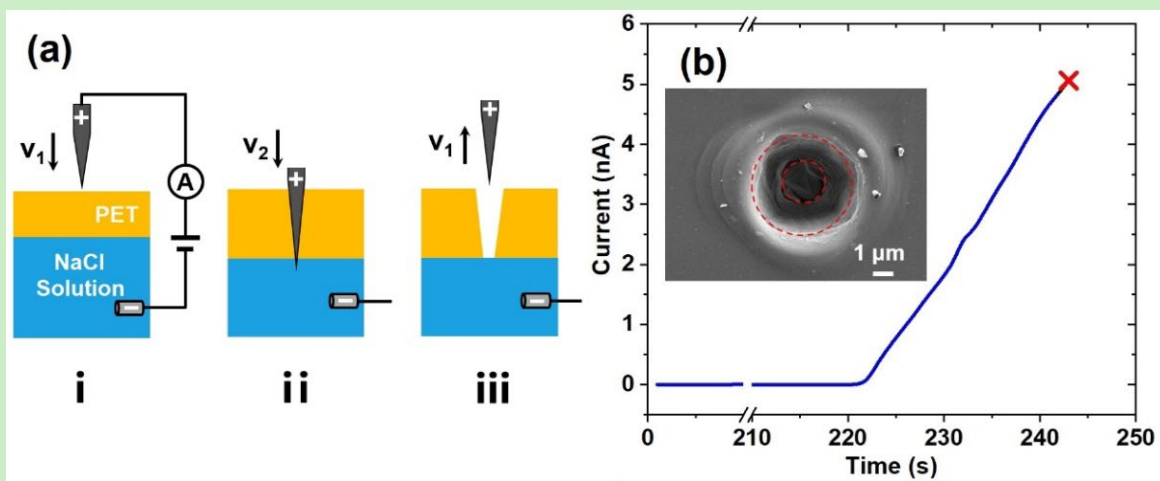

**Figure 2.** Fabrication process of conical micropores. (a) Scheme of micropore fabrication by the needle punching method. (b) Feedback current recorded in the process of a tungsten needle piercing a PET film. The inset shows the SEM image of the obtained conical micropore.

Based on the needle punching method, we propose two processes to fabricate micropores and nanopores, respectively. One is to feed the tungsten needle to pierce through the PET film to obtain conical micropores. The other is to form conical defects by feeding the tungsten needle piercing into the plastic film, and then obtain conical nanopores by chemical etching through the film.

Figure 2a shows the procedure for the preparation of conical micropores. A PET film is fixed on the reservoir, which is in contact with the NaCl solution on one side. The Ag/AgCl electrode is inserted into the reservoir. The tungsten needle is controlled to approach the PET film at a high speed of $v_1 = 1$ mm/s using a syringe pump and pauses at ~100 µm above the PET surface (Figure 2a i). Then, the needle continues feeding at a low speed of $v_2 = 1$ µm/s to pierce through the film to form a micropore (Figure 2a ii). After completing the pore preparation, the tungsten needle retracts at a high speed (Figure 2a iii). This fabrication process by the needle punching method is monitored by an electronic feedback circuit.

Figure 2b shows a current trace example obtained during the preparation of conical micropores. NaCl solution with a concentration of 1 M is selected. After the voltage of 1 V is applied, the current is 0 nA and remains constant, indicating that the tungsten needle has not penetrated through the PET film. At t=220 s, the current increases sharply, indicating that the tungsten needle has penetrated through the film and its tip is in contact with the solution. At this point, the micropore is formed. Subsequently, the tungsten needle continues to penetrate deeper into the solution, which further increases the pore size. When the feedback current reaches the threshold of 5 nA, the tungsten needle stops feeding and retracts in the reverse direction. The inset of Figure 2b is an SEM image of the obtained micropore. In the image, the contours of the tip and base of the conical pore (marked with red circles) can be distinguished. The measured diameters of the base and tip are 5.3 and 2.1 µm, respectively. The pore walls are smooth and the cone angle is ~15 °. More SEM images are shown in Figure S5. For different polymer films with various elastic moduli, the pore shape may become irregular.

## C. Controllable Preparation of Micropores

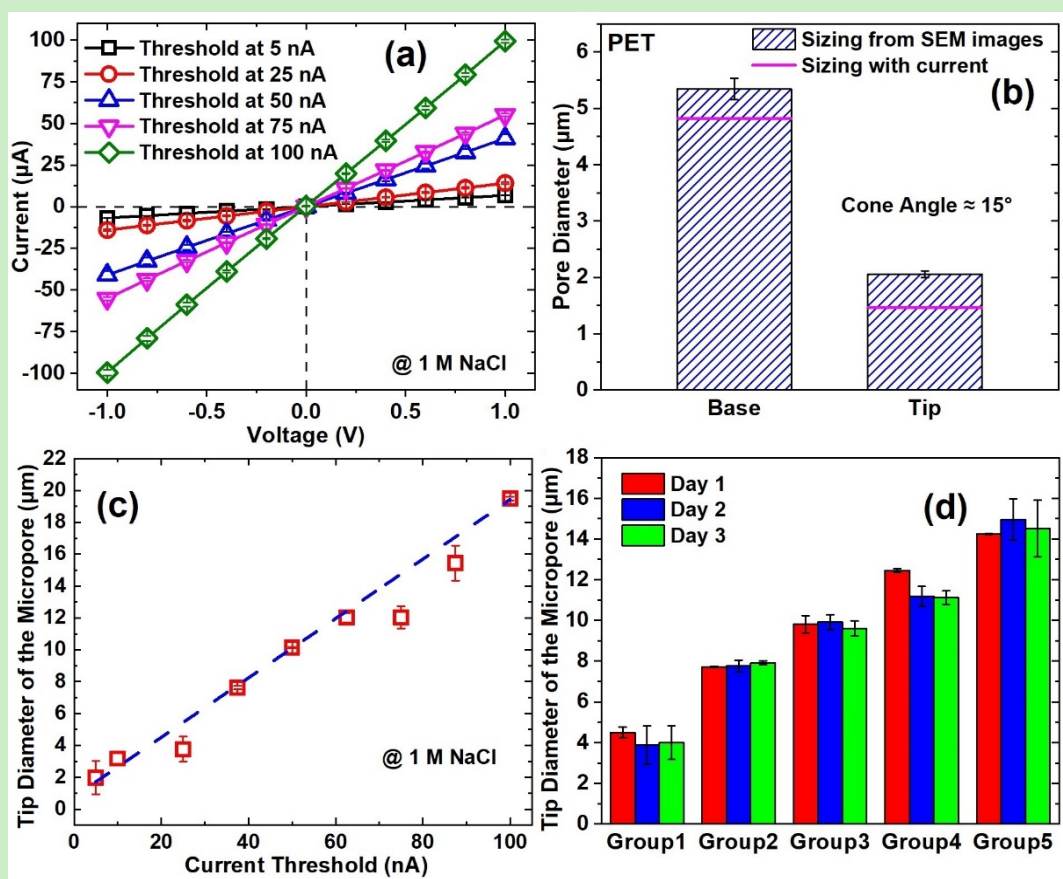

**Figure 3.** Controllable fabrication process of conical micropores with various sizes. (a) The current-voltage (I-V) curves in micropores prepared under different current thresholds. (b) The micropore size obtained by SEM imaging and evaluated through the I-V curve. (c) The distribution of tip diameter corresponding to different current

thresholds. (d) Stability of micropores.

Micropores of different sizes can be prepared by adjusting the current threshold.

Using micropores, the I-V curve can be obtained. The pore resistance $R$ is the inverse

of the I-V curve slope. Then, the tip and base diameters can be estimated with

equations 1 and 2.[28, 39, 55]

$$R = \frac{4L}{\pi \kappa dD} + \frac{1}{2\kappa d} + \frac{1}{2\kappa D} \tag{1}$$

$$D = d + 2L \tan(\pi\theta/360) \tag{2}$$

where, $d$, $D$, $L$, $\kappa$, and $\theta$ represent the tip diameter, base diameter, pore length,

conductivity of solution, and cone angle, respectively. In our case, $L$ = 12.5 µm and $\kappa$

= 8.33 S/m. PET is a kind of plastic material, so the micropore structure is relatively

stable. From the obtained SEM images, it can be seen that the cone angle of the

micropore is almost the same as that of the tungsten needle. Equations 1 and 2

provide a fast way to characterize the conical micropores without electron microscopy.

Using tungsten needles with a cone angle of 15°, micropores with different sizes

are successfully fabricated by controlling the current threshold. Figure 3a shows the

I-V curves through the conical micropores obtained under a threshold current varying from 5 nA to 100 nA. In 1 M NaCl solution, due to the large pore size and high concentration, ion current rectification (ICR) does not occur.[48] The I-V curves in the micropores prepared at different current thresholds have a linear profile, which displays excellent response characteristics. As the current threshold increases, the I-V curve in the obtained micropore has a larger slope, i.e. a larger pore size.[39, 48] The tip and base diameters of micropores can be calculated with equations 1 and 2. Figure 3b shows the comparison of conical pore sizes obtained from current predictions and SEM measurements. The SEM measurement provides a slightly larger dimension than that predicted by the equations. This may be due to the weakly hydrophobic surface of unetched PET surfaces. The micropore wall can form voids that affect the distribution of the solution, which increases the resistance of the micropore slightly.[56, 57] In general, the micropore sizes characterized by the two methods are similar with a deviation of 9.7%. It indicates that the size of micropores in the manufacturing process is determined by the shape of the tungsten needle tip. Note that the micropores prepared by the needle punching method can be chemically etched to remove a small amount of material from the surface.[48] The carboxyl groups (−COOH) can be formed on the surface[26] to change surface hydrophobicity and can also provide the basis for various chemical modifications.[5]

With prepared conical micropores under different thresholds and their sizes characterized by ion conductance through pores, the relationship between the predicted micropore sizes and the current thresholds is obtained, as shown in Figure 3c. The tip diameter of micropores is positively correlated with the current threshold. Therefore, according to the dependence of the pore tip diameter with the current threshold, the preparation of micropores with desired sizes can be realized. The procedure is simple to operate and highly reproducible.

When a tungsten needle punctures through the PET film to form micropores, the process will make the plastic film conduct elastic and plastic deformation. Considering that partial recovery of the micropore may occur after the removal of the external force, we characterize the stability of the fabricated conical micropores. After the preparation of micropores, they are preserved in deionized water. The I-V of the micropores is measured every 24 hours to calculate the pore size. As shown in Figure 3d and Figure S6, micropores with different tip diameters have high stability, and the shrinkage of their tip diameter is less than 10%.

### D. Nanoparticle Detection with a Micropore

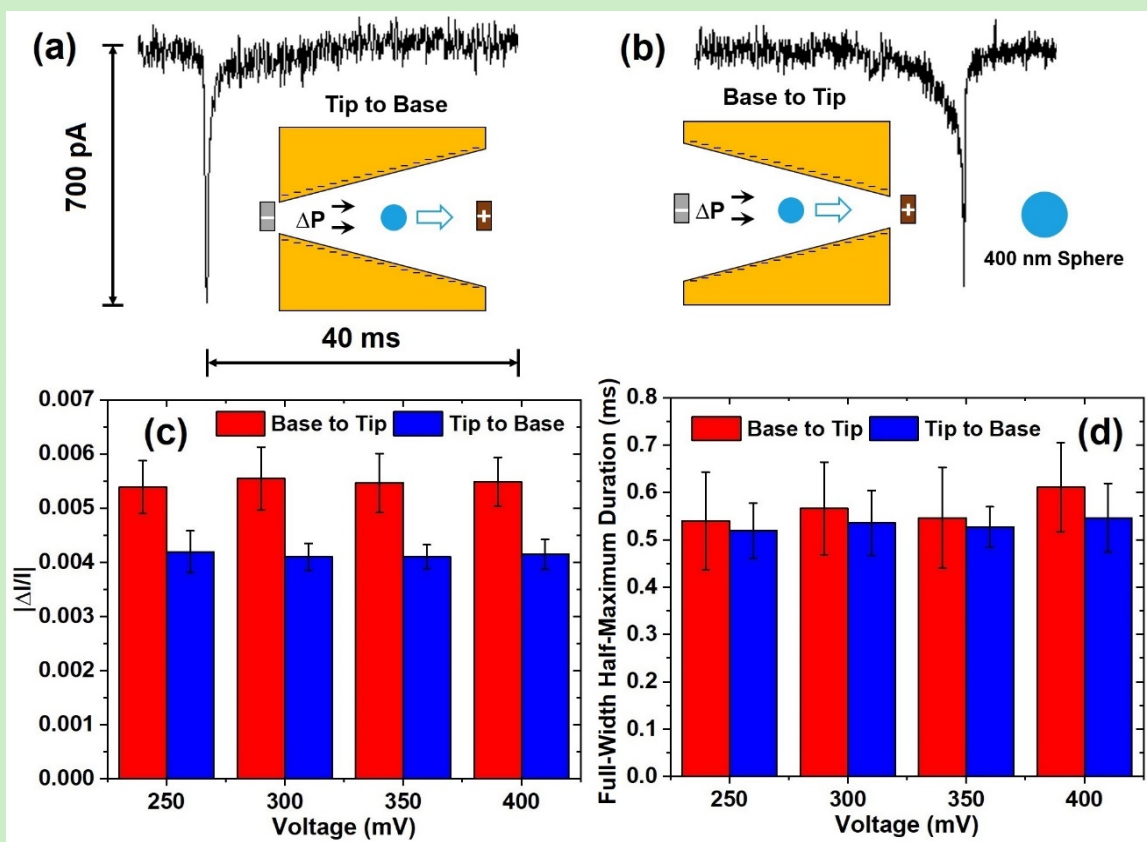

**Figure 4.** Particle detection using conical micropores. Example pulses and schematic diagram of the particle passing through the pore (a)-(b). (a)From tip to base. (b) From base to tip. (c) $|\Delta I/I|= |(I_b\text{-}I_0)/I_0|$ at different voltages. (d) Full-width half-maximum duration of the pulses. During the particle detection, a hydrostatic pressure (450 Pa) was applied on the particle side.

With the needle punching method, micropores can be prepared with different tip diameters ranging from 1 μm to tens of μm. These micropores can be used for nanoparticle detection based on the resistive-pulse technique. We choose a micropore with a cone angle of 15° and a tip diameter of 1173 nm for the detection experiments (Figure S7). The micropore is etched in 0.5 M NaOH solution at 70°C to form carboxyl groups to improve the hydrophilicity on pore walls.[25, 49] As shown in Figures 4a and 4b, the particle solution is added to the reservoir on one side of the film, and the KCl solution without nanoparticles is placed on the other side. Due to the weak surface charge density of the particle, the particles can not be moved by electrophoresis due to the large electroosmotic flow inside the conical pore. By adding a hydrostatic pressure on the particle side, the pressure difference can drive nanoparticles to pass through the pore. The change of ion current caused by the particle translocation through the micropore can be detected by applying a voltage across the pore. Figures S8 and S9 show the current traces recorded when particles pass through the conical pore in both directions, i.e. from the tip and the base, respectively.

Current blockades obtained when the nanoparticles enter the conical pore from opposite directions are different due to the asymmetric geometry of the conical pore.[25, 26, 48, 49] When the nanoparticle enters the pore from the tip, the current blockade signal first drops to a peak ($I_b$), and then slowly increases to the open current value ($I_0$)(Figure 4a). On the contrary, the current signal gradually decreases to a peak, and then sharply increases to the open current (Figure 4b). The resistive pulses of particles passing through micropores in both directions are analyzed to obtain the blockade current ratio $|\Delta I//I| = |(I_b-I_0)/I_0|$, and the full-width half-maximum duration corresponding to the duration time of the translocation.[11] As shown in Figure 4c, the $|\Delta I//I|$ is independent of applied voltage.[25] While the $|\Delta I//I|$ for the particle moving from the base to the tip is higher than that in the opposite direction. The directional dependence of $|\Delta I//I|$ with the conical pore is due to the regulation of ion concentration inside the pore when the particle appears at the tip.[48] Figure 4d shows the full-width half-maximum duration of the particle entering the conical pore in both directions at different voltages. Since the nanoparticles are mainly driven through the micropores by the applied hydrostatic pressure, the velocity of the particles is almost the same at different voltages.

**E. Fabrication of Nanopores**

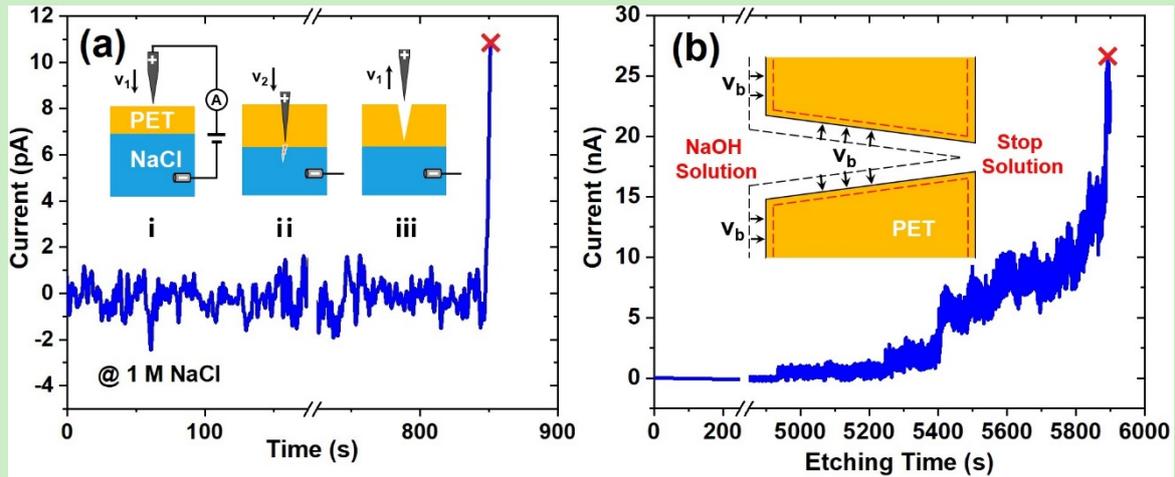

**Figure 5.** Fabrication process of conical nanopores. (a) Feedback current trace during the formation of conical defects. Insets show the formation scheme of a conical defect in the membrane. (b) Current trace example during the chemical etching process. The illustration shows the etching principle. $V_b$ represents the bulk etching speed of PET in NaOH solutions.

The size and shape of the tungsten needle tip affect the fabrication of conical micropores. Moreover, the nanoscale tip has a low mechanical strength which is prone to wear when the needle punches into the material. Therefore, it is extremely difficult to fabricate conical nanopores with a nanoscale tip diameter by punching through the membrane. Here, we propose a two-step fabrication method to prepare nanopores successfully. First, the tungsten needle is controlled to feed rapidly to the film surface under the observation with a microscope (Figure 5a). Then, the tungsten needle slowly pierces into the film at a rate of 100 nm/s. The current threshold of the feedback circuit is set to a very low value of 10 pA under 1 V. When the tungsten needle is about to pierce through the PET film, the leakage current is generated.[58]

With the current reaching the threshold, the tungsten needle stops and quickly retracts (Figure 5a). In this way, a conical defect is obtained in the film. Then, the PET films are etched by wet etching at room temperature. The etching solution 9 M NaOH is placed on the side of PET films with a conical defect, and the stop solution 1 M KCl and HCOOH solution is put on the other side ( Figure 5b).[48] The PET film in contact with the etching solution will be uniformly etched and its thickness will gradually decrease until a nanoscale pore is formed at the tip of the conical defect. After that, the stop solution reacts chemically with the etching solution to slow down the etching process.[3] This process is similar to the fabrication process of conical nanopores by the track-etching method. Figure 5b shows the current trace during the etching process. When the current is 0 nA, no conical nanopore is formed. With the etching lasting for ~5,000 s, the ionic current increases sharply, which indicates that a conical nanopore forms. As the current continues to increase, the size of the nanopore will further expand.[48] Note that during the chemical etching process, the film thickness is reduced which can be calculated accurately with the known bulk etching speed ($v_b$) of PET.[48]

**F. ICR Experiment with Conical Nanopores**

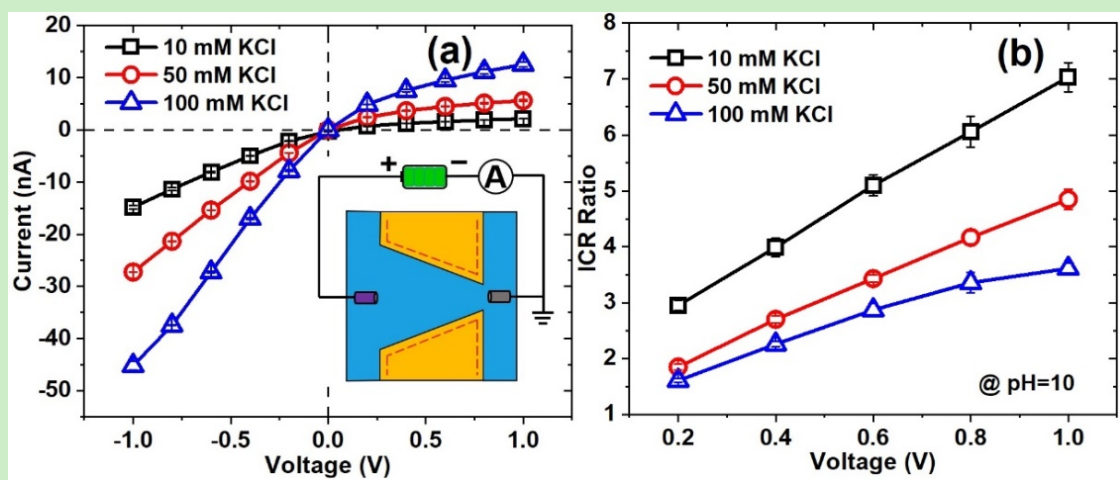

**Figure 6.** ICR in a conical nanopore. (a) ICR at different salt concentrations. (b) ICR ratio at different salt concentrations. The tip diameter, length, and cone angle of the conical pore are 79 nm, ~12.5 μm, and 15°, respectively.

A conical nanopore with a tip diameter of 79 nm was prepared on a 12.5-μm-thick PET film (Figure S10). Due to the asymmetric geometry of the conical nanopores, the ionic current rectification can be generated.[59, 60] Therefore, they have potential applications in ion circuits and current amplifiers.[5, 61] Because the ion transport in the nanopore has directional specificity, different ionic current is obtained under the same voltage but opposite polarities. When counterions move from tip to base, the ion current in the nanopore is larger, which is referred to as the "on" state. Conversely, the current inside the nanopore is smaller, which is known as the "off" state. The ICR phenomenon is well studied with experiments and simulations,[50, 62-64] which is mainly due to ion enrichment and depletion inside the pore caused by the ion selectivity of the tip.[59, 65, 66] When the ions in the nanopore are enriched or depleted, a higher or lower current value is obtained.

Using conical nanopores prepared by our two-step method, we obtain rectified I-V curves in 10-100 mM KCl solutions. As shown in Figure 6a, the nanopores show obvious ICR. At different concentrations, the ionic current at negative voltages is much larger than that at positive voltages. The ICR ratio at different voltages is shown in Figure 6b. With a constant concentration, the ICR ratio correlates positively with the voltage. Meanwhile, the ICR ratio gradually increases with the decrease of KCl concentration. Note that the ICR phenomenon within conical nanopores has been extensively investigated.[50, 62-64] We have not attempted to report the ICR within conical nanopores again. Our data shown here are used to characterize the properties of nanopores prepared by our fabrication method. From the obvious ICR phenomena, our nanopore surface is negatively charged.[59] From the structure of the PET material, negative surface charges are induced by the deprotonation of −COOH on nanopore walls. Further, the nanopores fabricated by our proposed method can be subjected to surface chemical modification, which can be widely used for the fabrication of functional nanopores.[5]

After conical nanopores are obtained, the size of the pore can be further uniformly enlarged by bulk etching of the PET material. As demonstrated in our previous work,[25] conical micropores can be obtained for nanoparticle detection by bulk etching a conical nanopore.

## IV. CONCLUSIONS

A low-cost and convenient method for preparing conical micropores based on the needle punching method is proposed. By controlling the tungsten needle to pierce through the polymer film, a micropore is formed, whose size can be monitored by the current feedback system. Based on the positive correlation between the tip diameter and the applied current threshold, the preparation of micropores with desired sizes can be realized. With the combination of needle punching and chemical etching, the preparation of nanopores can be achieved. Then, the fabricated micropore and nanopore are successfully used for nanoparticle detection with the resistive-pulse technique and ionic current rectification. From our results, the micro and nanopores prepared with our developed method exhibit a good sensing performance compared to those polymer pores obtained with the track-etching technique. Due to the advantages of low cost and simple operation, we think our fabrication method can provide a new solution for solid-state pore fabrication and accelerate the application of micro/nanopores in nanofluidic sensing.

**SUPPLEMENTARY MATERIALS**

The preparation process of the tungsten needle, the current signal of particle detection, and the size of micro/nanopores characterized in the supporting material.

**ACKNOWLEDGMENT**

This research was supported by the National Natural Science Foundation of China (52105579), the Natural Science Foundation of Shandong Province

## AUTHOR INFORMATION

### Author Contributions

**R. L.** and **Z. L.** contributed equally.

**Rui Liu:** Experimental design, experimental execution, data curation, initial draft writing. **Zhe Liu:** Experiment design, experiment execution, data curation, manuscript revision. **Jianfeng Li:** Carried out the review and revision of the draft. **Yinghua Qiu:** Conceptualization, methodology, resources, writing - original draft, review & editing, supervision, funding acquisition.

### Conflict of Interest

The authors declare no competing financial interest.